\documentclass[12pt,preprint]{aastex}
\usepackage{natbib}

\renewcommand{\ss}{\scriptsize}
\newcommand{\smhz}{\ensuremath{\mbox{\ss\ MHz}}}
\newcommand{\sghz}{\ensuremath{\mbox{\ss\ GHz}}}
\newcommand{\um}{\ensuremath{\micron}}
\newcommand{\expnt}[2]{\ensuremath{#1 \times 10^{#2}}}   
\newcommand{\gsim}{\gtrsim}

\newcommand{\mc}{\multicolumn}
\newcommand{\ujy}{\ensuremath{\mbox{ }\mu\mbox{Jy}}}

\newcommand{\sgr}{SGR~1900$+$14}
\newcommand{\psr}{PSR~J1907$+$0918}
\newcommand{\snr}{G042.8$+$00.6}
\newcommand{\nsnr}{G043.5$+$00.6}
\newcommand{\err}{\ensuremath{0\farcs79}}
\newcommand{\stislim}{29.0}
\newcommand{\klim}{20.8}
\newcommand{\jlim}{22.8}

\shorttitle{Multi-Wavelength Observations of \sgr}
\shortauthors{Kaplan, Kulkarni, Frail, \& van~Kerkwijk}

\begin{document}
\title{Deep Radio, Optical, and Infrared Observations of \sgr}
\author{D. L. Kaplan, S. R. Kulkarni}
\affil{Department of Astronomy, 105-24 California Institute of
Technology, Pasadena, California 91125, USA}
\email{dlk@astro.caltech.edu, srk@astro.caltech.edu}
\author{D. A. Frail}
\affil{National Radio Astronomy Observatory, Socorro, NM 87801, USA}
\email{dfrail@nrao.edu}
\and \author{M. H. van Kerkwijk}
\affil{Sterrenkundig Instituut, Universiteit Utrecht, Postbus 80000,
3508 TA Utrecht, The Netherlands}
\email{M.H.vanKerkwijk@phys.uu.nl}
\slugcomment{Accepted by ApJ}

\begin{abstract}
We present {\em HST}/STIS, Keck $J$- and $K_{s}$-band, and VLA 332-MHz
and 1400-MHz images of the region around the Soft $\gamma$-ray
Repeater \sgr.  No non-stellar sources
were detected at the position of \sgr, giving 
3-$\sigma$ limits of $S_{332\smhz}  < 6$~mJy, $S_{1.4\sghz}
 < 1.5$~mJy, $m_{\rm 50CCD}\gsim \stislim$~mag, $J \gsim \jlim$~mag, 
$K_{s} \gsim \klim$~mag (point sources), and 
$S_{332\smhz} < 6.1\mbox{ mJy arcmin}^{-2}$, $S_{1.4\sghz} <
6.2\mbox{ mJy arcmin}^{-2}$ (extended emission).  Given the very high
extinction, the STIS and $J$-band non-detections are not constraining,
but the $K_{s}$-band limit rules out basic accretion disk models for
\sgr, and is almost comparable in depth to the $K_{s}$-band detection for the
anomalous X-ray pulsar
4U~0142$+$61.  Finally, we report the detection in this field of
three new 
candidate supernova remnants, SNRs \nsnr, G042.0$-$00.1,  and G041.5+00.4.
\end{abstract}


\section{Introduction}
Soft $\gamma$-ray Repeaters (SGRs; see \citealt{h99} for a recent
observational review) are sources both of bursts of
hard X-rays/$\gamma$-rays and softer quiescent X-ray emission.  They
are generally thought to be
young ($<10^{4}$~yr) neutron stars with extremely strong magnetic
fields, or magnetars \citep[e.g.][]{dt92,td93}.  This belief was 
motivated by the energetics of their bursts \citep{p92,td95},
their associations with young supernova remnants (SNRs;
 \citealt{ekl+80,kf93,vkfg94}), and the
detection of 
X-ray pulsations with long (5--10~s) periods and large ($\sim
10^{-11}\mbox{ s s}^{-1}$) spin-down rates \citep{kds+98,hlk+99}.
However, more recent observational evidence
\citep{lx00,hkc+99,kfk+02} has shown that in most cases the
SGR/SNR associations are false and can be attributed to random chance
\citep{gsgv01}.  Instead of being constrained by the
SNR associations, the small ages of the SGRs are now limited by the
proximity of the SGRs to sites of very massive star formation
\citep{fmc+99,vhl+00}. 

An alternate theory, first proposed \citep*{vptvdh95,chn00}
for the emission  
from the related  Anomalous X-ray Pulsars (AXPs; see \citealt{m99}),
and now for 
the SGRs as well (\citealt{mlrh00,mlrh99}; but see \citealt{d01}), is
accretion from a 
``fallback'' disk of material produced during the supernova
explosion onto a central neutron star. 
In this model, the neutron stars are spinning near the equilibrium
period which is proportional to the dipole field strength.  This model
then explains the spin-down and period clustering of the AXPs:
for a typical field strength of $10^{12}$~G, the resultant spin period
is $\sim 8$~s.

To date, no confirmed persistent counterparts have been found for SGRs at
optical/IR or radio wavelengths \citep[e.g.][]{kkvk+01}, although
\citet{eik01} may have found an infrared counterpart to SGR~1806$-$20.
It generally has been the case that when counterparts are identified
at other wavelengths, especially optical or IR, there is considerable
progress in our understanding of these objects.  For example, the
identification of an optical/IR counterpart of the AXP 4U~0142+61
\citep*{hvkk00} has posed the strongest challenge to accretion models.
In view of this, it is important to press upon sensitive
multi-wavelength observations of SGRs.  Here we present optical, IR,
and radio observations of the very accurately localized \sgr\
(\citealt*{fkb99}; \citealt{hlk+99}).

\subsection{\sgr}
\sgr\ had been a relatively innocuous source until late 1998.    The
source of sporadic bursting emission (\citealt*{mgg79};
\citealt{kfm+93}), it had 
been associated with a soft, persistent X-ray source \citep{hlv+96}
and a SNR, \snr\
\citep[][]{vkfg94}.  On 1998~August~27, the spacecraft Konus detected
an intense burst of $\gamma$-rays \citep{hcm+99}.  Arrival-time
localization \citep{hkw+99b} 
soon identified \sgr\ as the 
source.  Prompt radio observations detected a fading, non-thermal
source \citep{fkb99} coincident with the X-ray source \citep{hlk+99}, and timing of
the X-ray emission revealed a 5.16-s pulsar \citep{hlk+99}
 with magnetar-like
spin-down \citep{ksh+99}.
The radio source associated with the 1998~August~27 burst 
is located at  $\alpha_{2000}=19^{\rm h}07^{\rm m}14\fs33$,
$\delta_{2000}=+09\degr19\arcmin20\farcs1$, with uncertainty $\pm 0\farcs15$
in each coordinate \citep{fkb99}.  

Searches for \sgr\ at wavelengths other than X-ray and $\gamma$-ray
have not detected any  quiescent counterpart
\citep{hlv+96,lx00,ed00,vhl+00}.  However, \citet{lx00} did detect a
young ($10^{4-5}$~yr old)
radio pulsar (\psr) located in the vicinity\footnote{\psr\ is located at $\alpha=19^{\rm
h}07^{\rm m}22\fs441$, $\delta=+09\degr18\arcmin30\farcs76$, with
uncertainty $\pm 0\farcs05$ in each coordinate \citep{lx00}.} of both \sgr\ and \snr, suggesting
that \snr\ could be associated with \psr\ just as easily as with
\sgr.

\begin{figure*}
\plotone{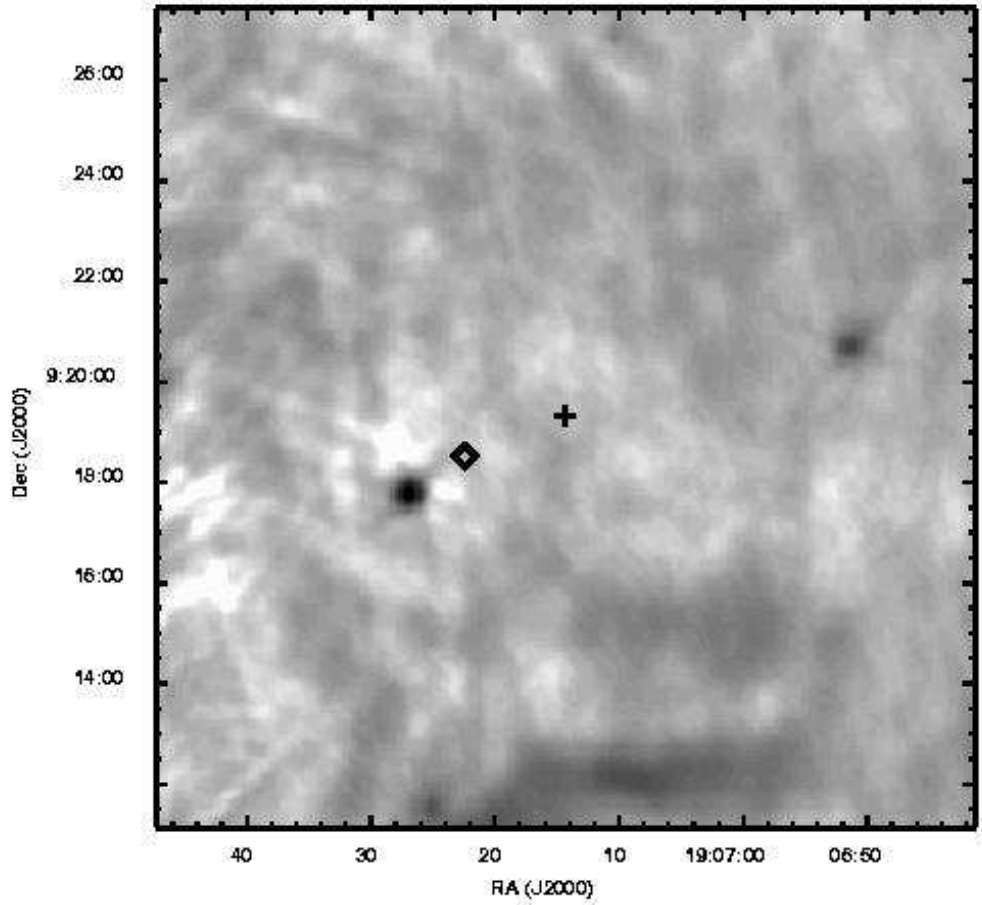}
\caption{1.4-GHz VLA image of the region around \sgr.  The cross
marks the position of \sgr, and the diamond marks the position of \psr.  The
northern rim of SNR \snr\ is visible.  The radio source to the
south-east of \psr\ is not related.
\label{fig:l}
}
\end{figure*}

\begin{figure*}
\plotone{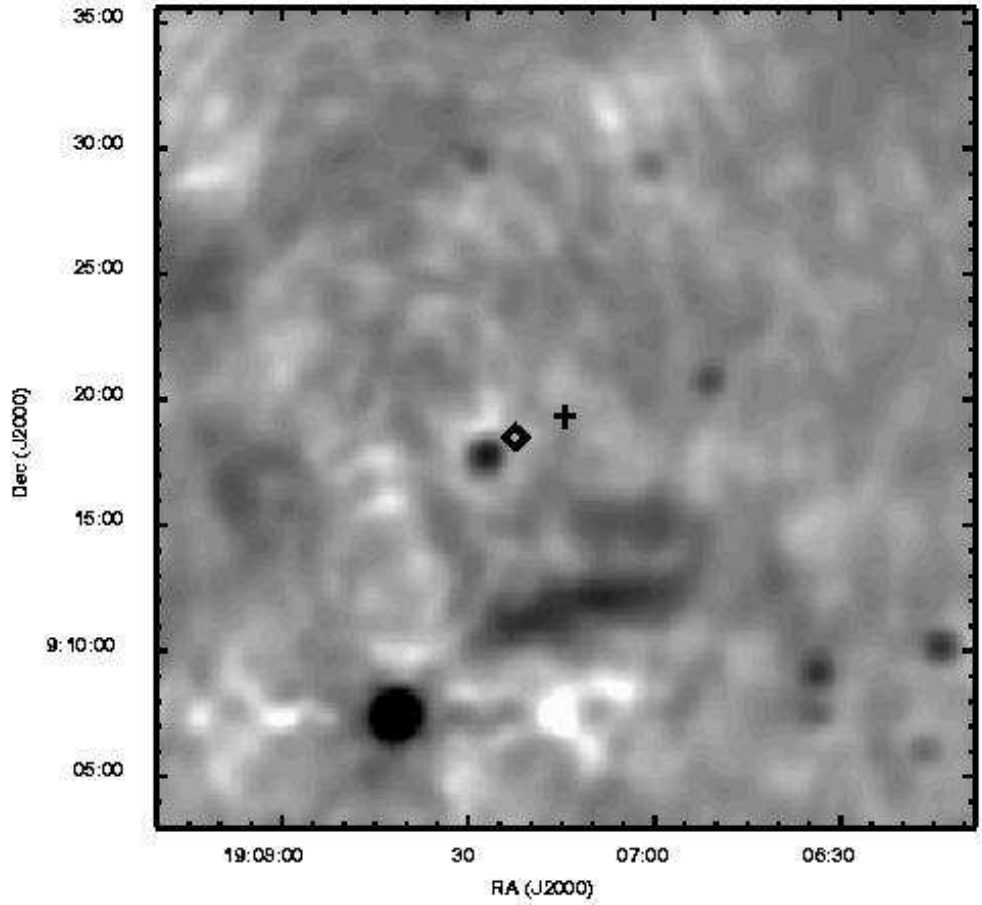}
\caption{1.4-GHz VLA image of the region around \sgr, tapered to $\sim
60\arcsec$ resolution.  The cross
marks the position of \sgr, and the diamond marks the position of
\psr.   The radio source to the south-east of \psr\ is not related. 
\label{fig:l-tap}
}
\end{figure*}

\citet{vhl+00} have suggested that \sgr\ is somehow related to a
compact cluster found in its vicinity (about $12\arcsec$ away).  A
similar cluster has been identified near SGR~1806$-$20 \citep{fmc+99}.
We note, though, that no such cluster is found in the vicinity of
SGR~0526-66 in the {\em HST} images presented by \citet{kkvk+01}.
Furthermore, SGR~1806$-$20 and \sgr\ are found at low Galactic
latitudes in the inner Galaxy and the chance coincidence probability
with star clusters, the rare Luminous Blue Variables (LBVs) and SNRs
is not negligible \citep[e.g.][]{gsgv01}.  Indeed, there is now significant
doubt about the association of SGR~1806$-$20 with the SNR G10.0$-$0.3
\citep{hkc+99,kfk+02}.  Nonetheless, the presence of star
clusters in the vicinity of \sgr\ appears intriguing, but while our
radio maps are 
deeper than previous images \citep[e.g.][]{vkfg94} we find no
radio source or nebulae associated with this star cluster (see
Figures~\ref{fig:l}--\ref{fig:p-wide}). 

For the X-ray properties of \sgr\ in what follows, we use the
power-law $+$ blackbody 
fit to the second epoch (the one closer to  normal quiescence) of the
recent {\em Chandra} observations 
by \citet{fkkf01}: $N_{H} = \expnt{2.3 \pm 1}{22}\mbox{ cm}^{-2}$,
power-law index 
$\Gamma =1.9 \pm 0.3$, blackbody temperature $kT \approx 0.49 \pm 0.03$~keV, and
an un-absorbed 0.5--10~keV X-ray flux 
$f_{\rm X}^{\rm U}=\expnt{2.0}{-11}\mbox{ erg s}^{-1}\mbox{ cm}^{-2}$
\citep[also see][]{ktw+01}.  This is consistent with the {\em ASCA}
measurements of \citet{hlk+99}.
We  adopt a distance $d=5$~kpc \citep{hlk+99}, which 
is based on hydrogen absorption.  The hydrogen column density implies
a visual extinction 
$A_{V}=12.8 \pm 0.8$~mag, assuming $N_{H} = \expnt{1.79 \pm 0.03}{21}A_{V}\mbox{
cm}^{-2}$ \citep{ps95}.  These values differ significantly from those assumed for the
star cluster ($d=12$--15~kpc, $A_{V}\approx 19.2$~mag;
\citealt{vlh+96}), and this difference may make the case for  associating \sgr\ and
the cluster less compelling.

\section{Observations}
We have carried out a campaign at optical/IR and radio wavelengths to
detect \sgr.  The radio data can also determine the existence of a low
surface-brightness nebula surrounding \sgr, and localize \sgr\
relative to nearby SNRs.  See Table~\ref{tab_obs} for a log of all the 
observations.

\subsection{Optical}
The optical data  consist of 5 exposures taken  by the Space
Telescope Imaging 
Spectrograph (STIS) aboard the {\em Hubble Space Telescope} ({\em HST}), totaling 5195~s.  The data
were taken in imaging mode using the clear (50CCD) filter, giving a
bandpass centered at roughly $585.2$~nm and a full-width at
half-maximum (FWHM) of $\approx 400$~nm.  The first exposure had a
duration of 30~s, and was used for astrometry (see below).  The
remaining exposures were $\sim 1300$-s long and were dithered in the usual
manner to improve resolution and mitigate the
effects of CCD irregularities.

The reduction followed the standard procedure for STIS images.  We
followed the standard drizzling reduction 
procedure \citep{fm98}
expanding the individual pixels by a factor of two.
This drizzling changed the pixel scale from $0\farcs05\mbox{
pixel}^{-1}$ to $0\farcs025\mbox{ pixel}^{-1}$.  Typical stellar
sources had a FWHM of $0\farcs095$; see Figure~\ref{fig:stis}.

For the astrometry, we selected from the USNO-A2.0 catalogue
\citep{m98} all 76 stars that overlapped with a 30-s R-band 
image taken on 28 August 1998 with the Low-Resolution Imaging
Spectrograph \citep[LRIS;][]{o+95} on the Keck telescope.  For
the 54 stars that were not overexposed, we measured centroids and
corrected for instrumental distortion using a bi-cubic function
determined by J.~Cohen (1997, private communication).  We used these
to solve for the zero-points in each coordinate, the plate scale, and
the position angle on the sky.  After rejecting 8 outliers (residual
larger than $0\farcs8$), the root-mean-square residuals were $0\farcs29$
in each coordinate.  This is somewhat larger than we found for other
projects, but not atypical for relatively crowded Galactic fields on
the Northern hemisphere.  Given this, the astrometry is tied to the
USNO-A2.0 system at the $0\farcs05$ level.  The typical accuracy with
which the USNO-A2.0 catalogue is on the International Celestial
Reference Frame (ICRF) is about $0\farcs2$ \citep{d99}.  

The astrometry was transferred to the STIS image using 43 
stars close to the radio position, solving again for plate scale,
rotation and zero-points.  The rms residuals were $0\farcs02$
in each coordinate, giving an overall astrometric uncertainty of
$0\farcs21$ in each coordinate.   The 3-$\sigma$
limiting magnitude, based on calibration using the photometric
keywords in the image 
header, is $\stislim$, or $f_{\nu,\rm min}= 10$~nJy
\footnote{Note that STIS magnitudes are in the STMAG system, where
$m=-21.1 -2.5 \log F_{\lambda}$, with $F_{\lambda}$ in units of
$\mbox{ erg s}^{-1}\mbox{ cm}^{-2}\mbox{ \AA}^{-1}$.  All other magnitudes
are Vega-based.}.


\subsection{Infrared}
The infrared data were taken with the Near Infrared
Camera \citep[NIRC;][]{ms94} mounted on the 10-m Keck~I
telescope.
The observations had photometric conditions and very good seeing ($0\farcs35$ at $K_{s}$-band).
The  data were reduced with custom {\tt IRAF}
software.  The images were dark subtracted, flat-fielded, and corrected
for bad pixels and cosmic rays.  We then made object masks, which were
used to construct improved flat fields for a second round of
reduction.  The data were finally registered, shifted, and co-added; see
Figure~\ref{fig:nirc}. 

The astrometry was performed with 50 stars from the deep STIS image.  This
solution had errors of $\pm0\farcs04$ (in each coordinate) relative to the STIS image, or
again $\pm0\farcs21$ relative to the ICRF.  For photometric calibration
we used the standard star SJ~9101 \citep{pmk+98}.  The 3-$\sigma$
limiting magnitudes are $K_{s} \approx \klim$ and $J\approx \jlim$,
corresponding to $f_{\nu, K_{s}} \approx 3.1\ujy$ and $f_{\nu, J} \approx
1.2 \ujy$, based on the calibration of \citet*{bcp98}.

\subsection{Radio}
\label{sec:radio}
The 1.4-GHz data were taken at the Very Large Array
(VLA) in its C and D configurations using  the standard
continuum mode, with $2\times 50$~MHz 
bandwidth.  The separate observations were calibrated independently, following the
standard procedure in {\tt AIPS}.   The phase calibrator was
JVAS~J1925+2106, and the flux calibrator was 3C~48.
The data were then self-calibrated and
combined for imaging.  The nominal beam size was $\approx 20\arcsec$,
and the rms noise is $\approx 0.5\mbox{ mJy
beam}^{-1}$; see Figure~\ref{fig:l}.

To aid in the detection of diffuse structure we
also imaged using a $uv$-taper of $2.5\mbox{ k}\lambda$, giving an effective
beam size of $\approx 60\arcsec$.  The rms noise is $\approx
2\mbox{ mJy beam}^{-1}$, largely due to  confusion from
unresolved sources in the Galactic plane; see Figure~\ref{fig:l-tap}.

\begin{figure*}
\plotone{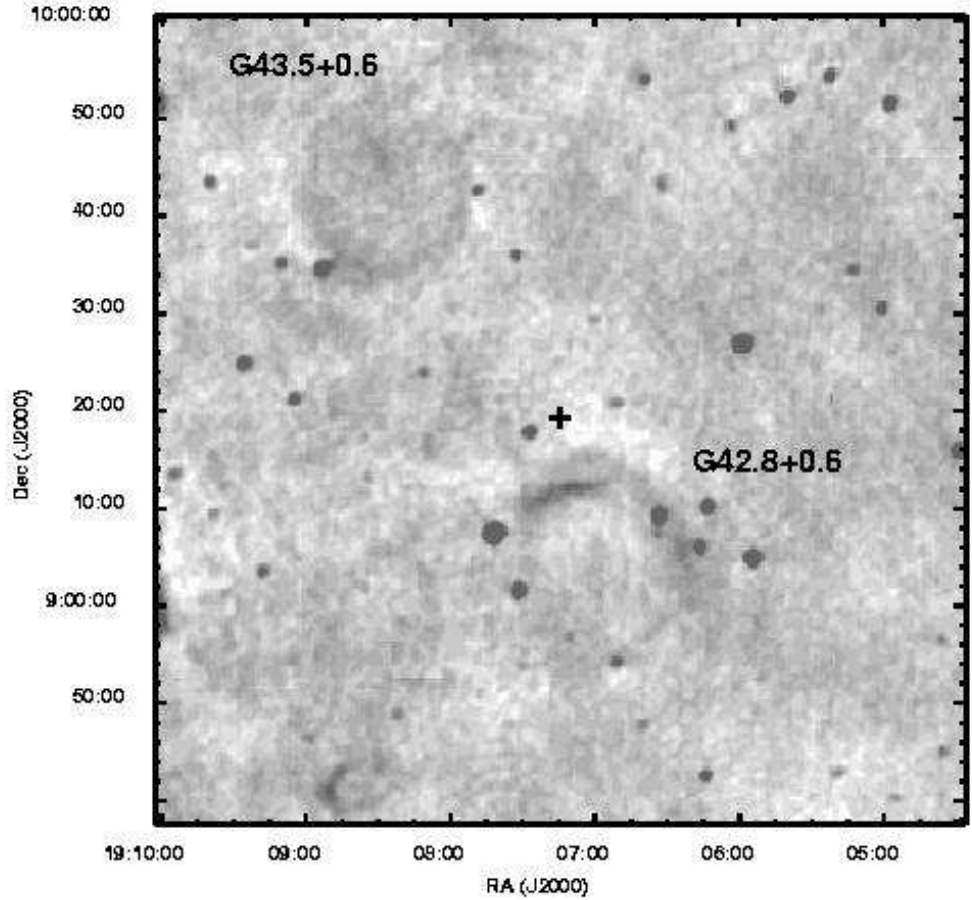}
\caption{332-MHz VLA image of the region around \sgr.   SNR \snr\ and
candidate SNR 
\nsnr\ are labeled.  The cross
marks the position of \sgr; there is no radio source at this position.
The white (i.e.\ 
negative) region near \sgr\ and \psr\ is an artifact of the
deconvolution process due to incomplete coverage of the $uv$-plane.
\label{fig:p-zoom}
}
\end{figure*}

The 332~MHz data were taken at the VLA in spectral-line mode with 32
channels of 97~kHz each to aid in the 
excision of radio-frequency interference (RFI).
We reduced the  observations    in {\tt AIPS}, following the procedure
for low-frequency data reduction\footnote{{\tt
http://rsd-www.nrl.navy.mil/7213/lazio/tutorial/index.html}}
\citep[see also][]{lkl+00}.  First we
removed any egregious RFI.  Then we  phase (using 3C~380) and
gain (using 3C~286 and 3C~48) calibrated the fringe visibilities.  We again searched for
and removed RFI, then averaged the 
channels together.  We  mapped and self-calibrated the data iteratively, using
 3-dimensional Fourier transforms in {\tt IMAGR}, the basic {\tt AIPS}
deconvolution and imaging task.  
While imaging we used 25 planar facets arranged
in a $5\times 5$ grid, each of which had $256 \times 256$ $13\arcsec$
cells.  These facets were used to mitigate the effects of non-coplanar
arrays that are present at low frequencies and with wide fields \citep{cp92,perl99}.  Finally,
we re-assembled the facets and flattened them into a single planar map
using the task {\tt 
FLATN}.  The rms noise is $\approx 2\mbox{ mJy beam}^{-1}$, limited
primarily by confusion due to unresolved emission in the Galactic
plane and  to sidelobes from the bright W49 complex.  The final
beam size is $\approx 50\arcsec$, comparable to that 
of the tapered 1.4-GHz images; see Figures~\ref{fig:p-wide} and \ref{fig:p-zoom}.

\section{Analysis \& Discussion}
\subsection{Optical/IR}
\begin{figure*}
\plotone{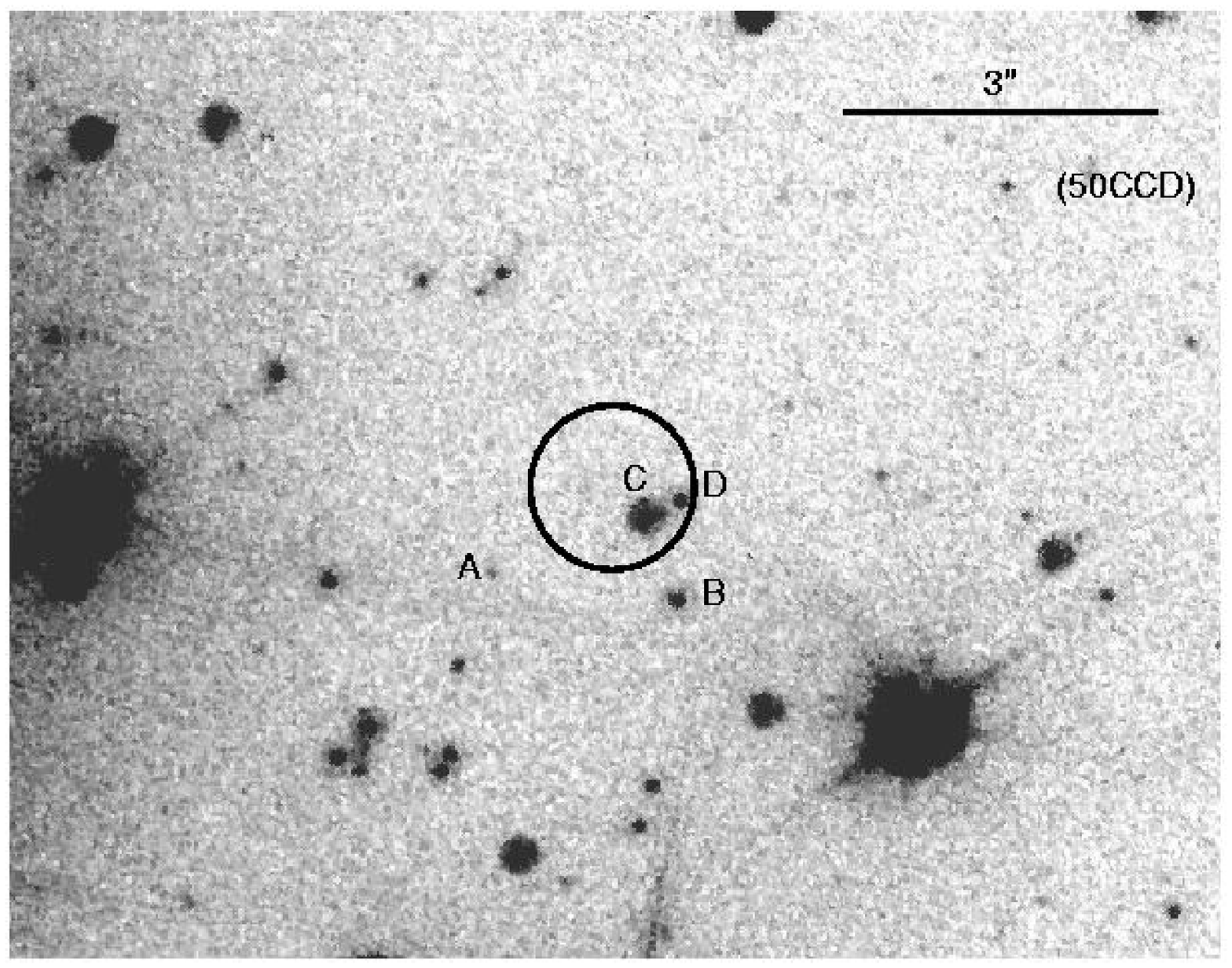}
\caption{{\em HST}/STIS image of the region around \sgr, showing the
\err-radius error circle around the position of the transient
radio source.  Sources A--D are indicated.  North is up, east is to
the left, and the scale bar indicates $3\arcsec$.
\label{fig:stis}
}
\end{figure*}

\begin{figure*}
\plottwo{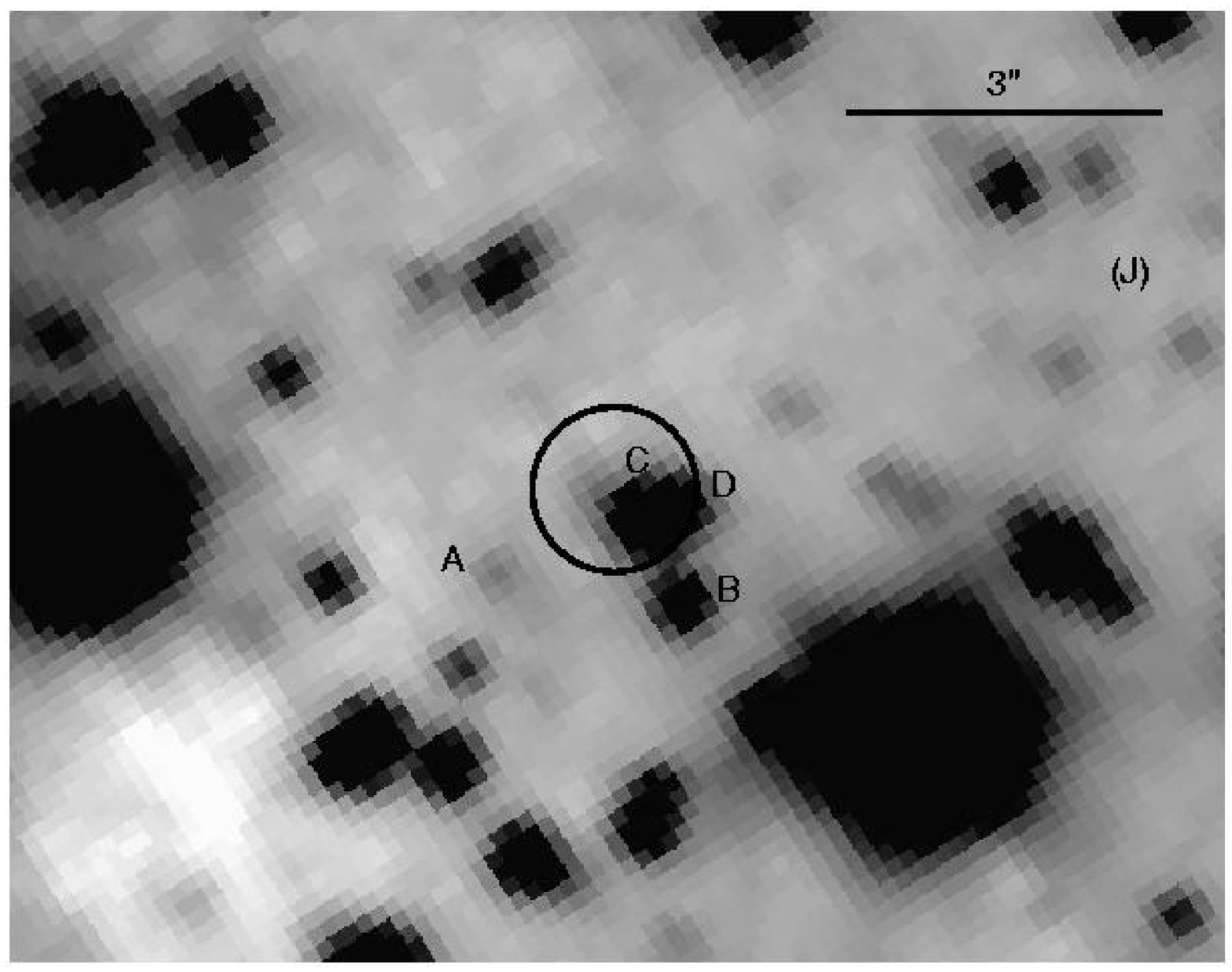}{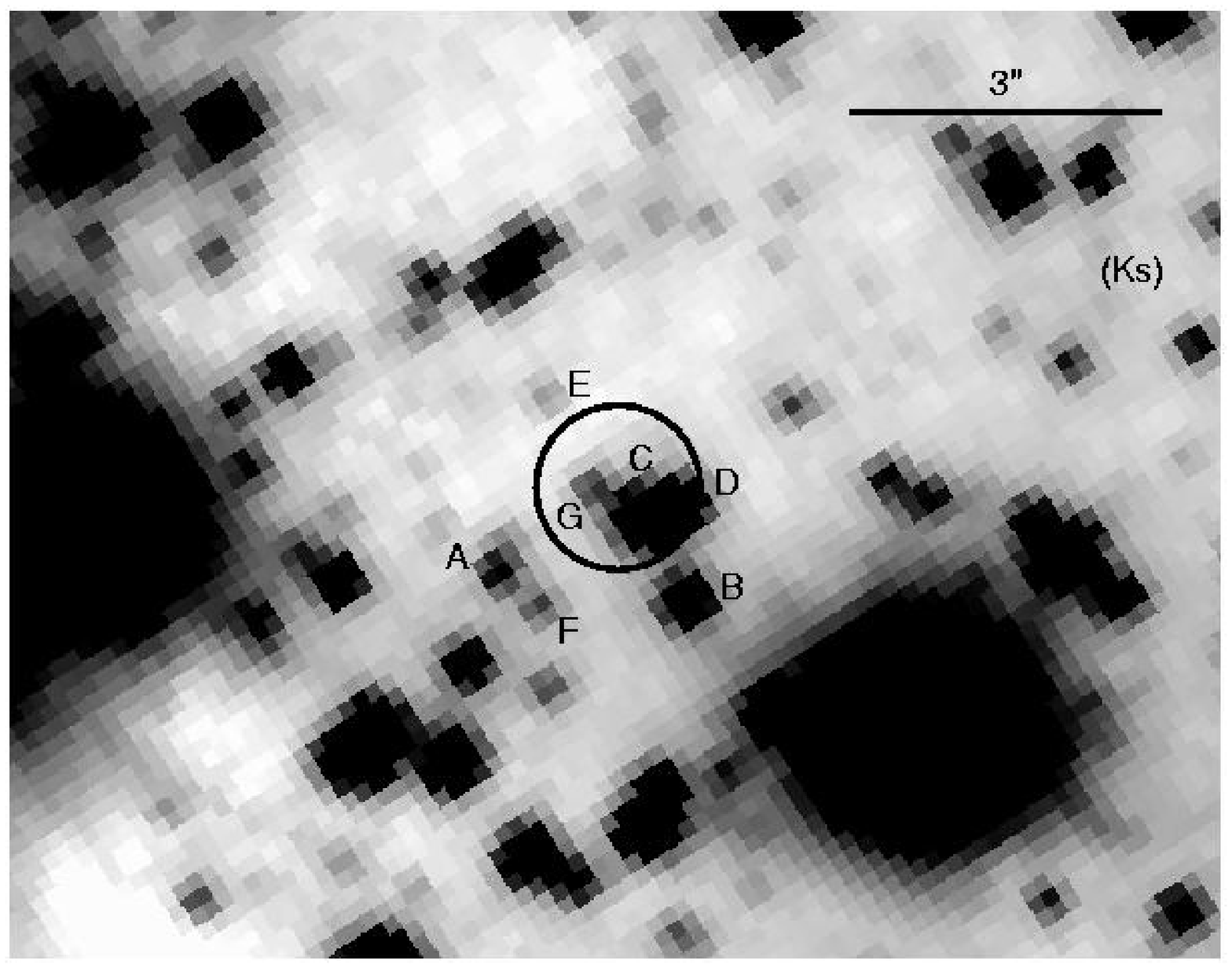}
\caption{NIRC images of the region around \sgr, showing the
\err-radius error circle around the position of the transient
radio source.  $J$-band (left); $K_{s}$-band (right).  Sources A--G
are indicated.  We were unable to perform photometry for source
G in any image.  North is up, east is to the left, and the scale bar
indicates $3\arcsec$. 
\label{fig:nirc}
}
\end{figure*}

For the infrared and optical observations the astrometric uncertainty is $0\farcs21$.
Combined with the $0\farcs15$ uncertainty from \citet{fkb99}, this
gives us a 99\% confidence radius of \err.  We examined all
sources in or near a circle of this radius around the position of
\citet{fkb99}.  In the deepest observation, the $K_{s}$-band image, we 
detect 7 sources that we label A--G on Figures~\ref{fig:stis} and
\ref{fig:nirc} (only sources A--D are visible in
Figure~\ref{fig:stis}).  Of these, C and D  
are significantly blended on all the images, as are sources C and G.
Sources E, F, and G are too faint to measure on the STIS image, and
sources E and F are too faint to measure on the $J$-band image.
Because of the degree of blending, we used a point-spread function
(psf) fitting and subtraction technique to perform photometry on the
infrared data, using
the {\tt DAOPHOT} package.  First we measured the psf's of a number
($\approx 30$) of stars in the field without close neighbors, and
constructed an average psf for each image.  We then fit this psf to
each source in Figures~\ref{fig:stis} and
\ref{fig:nirc},  subtracted the fitted stellar profile
from the images, and used the subtracted images to fit the sources
that were hidden by the original source.  After iterating this
procedure, we were able to obtain consistent magnitudes for most of
the sources.  For the sources that we were able to fit, all had sizes
consistent with unresolved sources.  For the STIS data, we adjusted
the measured magnitudes to a nominal infinite aperture with a $0.13
\pm 0.03$~mag correction, based on the observed STIS psf\footnote{{\tt
http://www.stsci.edu/instruments/stis/documents/ihb/stis\_cy11\_ihb.pdf}}.
We present the positions and
magnitudes of sources A--G in Table~\ref{tab:stars}.  

\begin{figure*}
\plotone{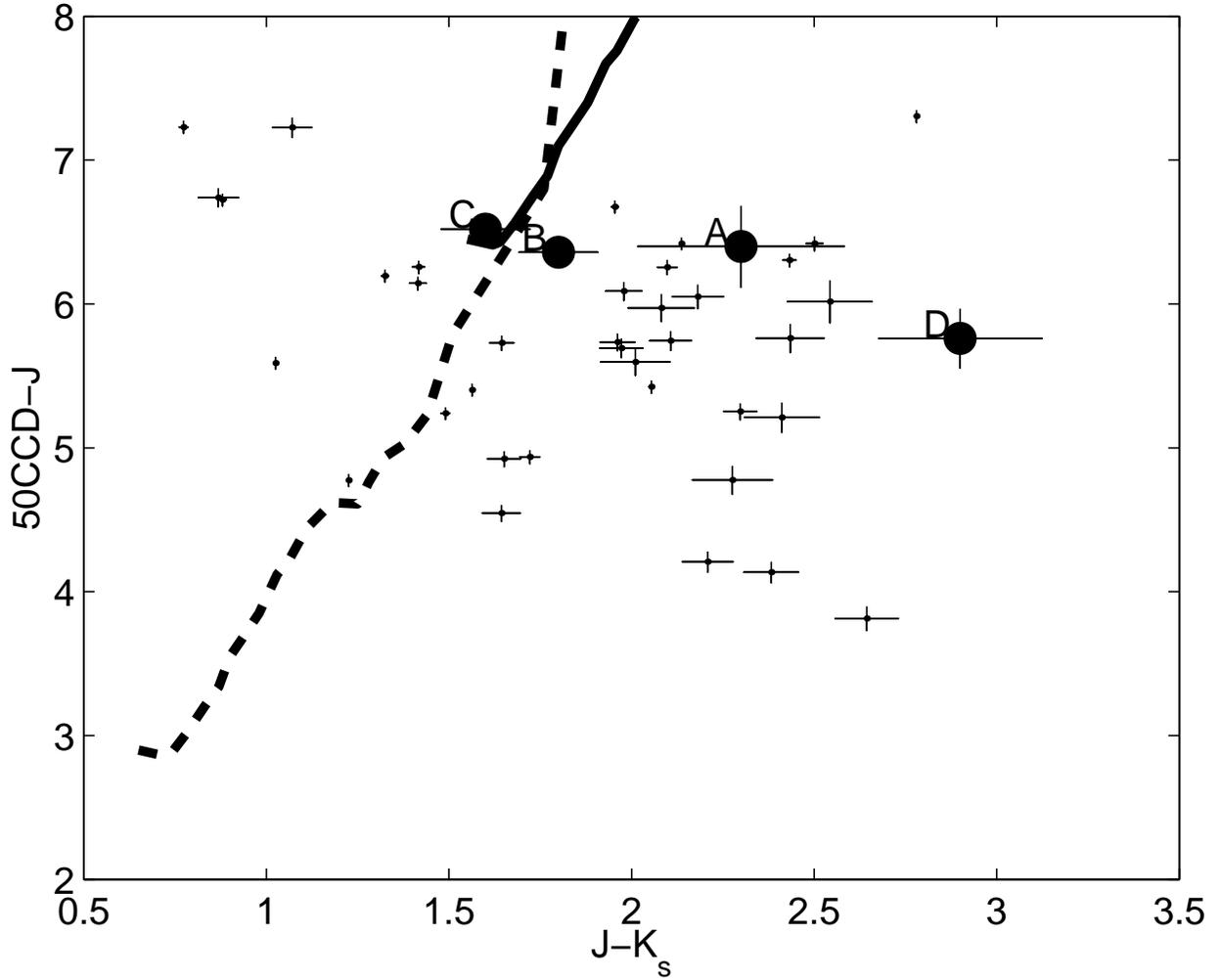}
\caption{$m_{\rm 50CCD}-J$ vs.\ $J-K_{s}$ color-color diagram for the sources in Table~\ref{tab:stars}
(circles) and 37 field stars (points).  Also plotted are model
main-sequence colors, assuming $V \approx
m_{\rm 50CCD}$, for $A_{V}=10$~mag (solid line) and $A_{V}=5$~mag
(dashed line), both at a distance of 8~kpc.  } 
\label{fig:cc}
\end{figure*}

\begin{figure*}
\plottwo{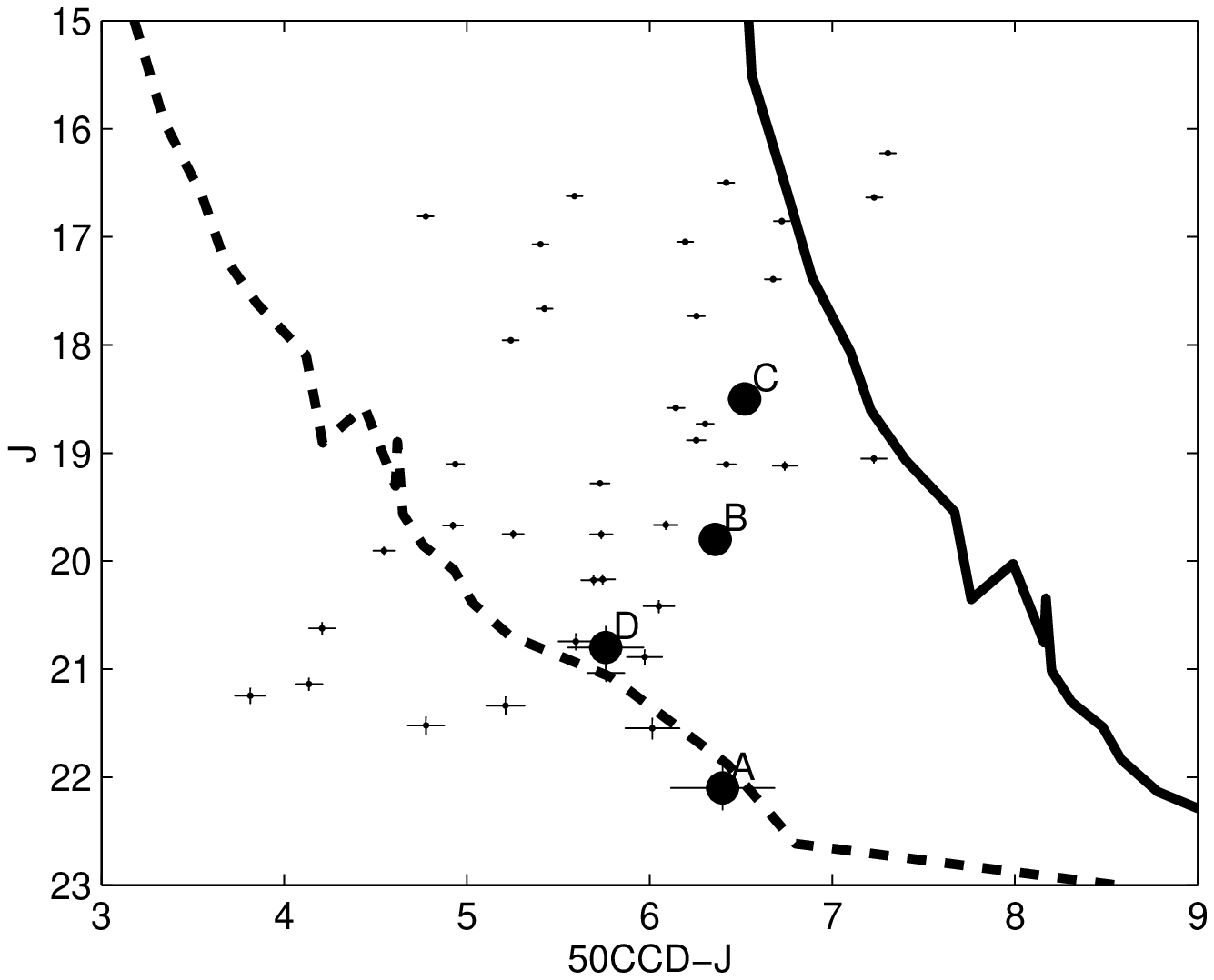}{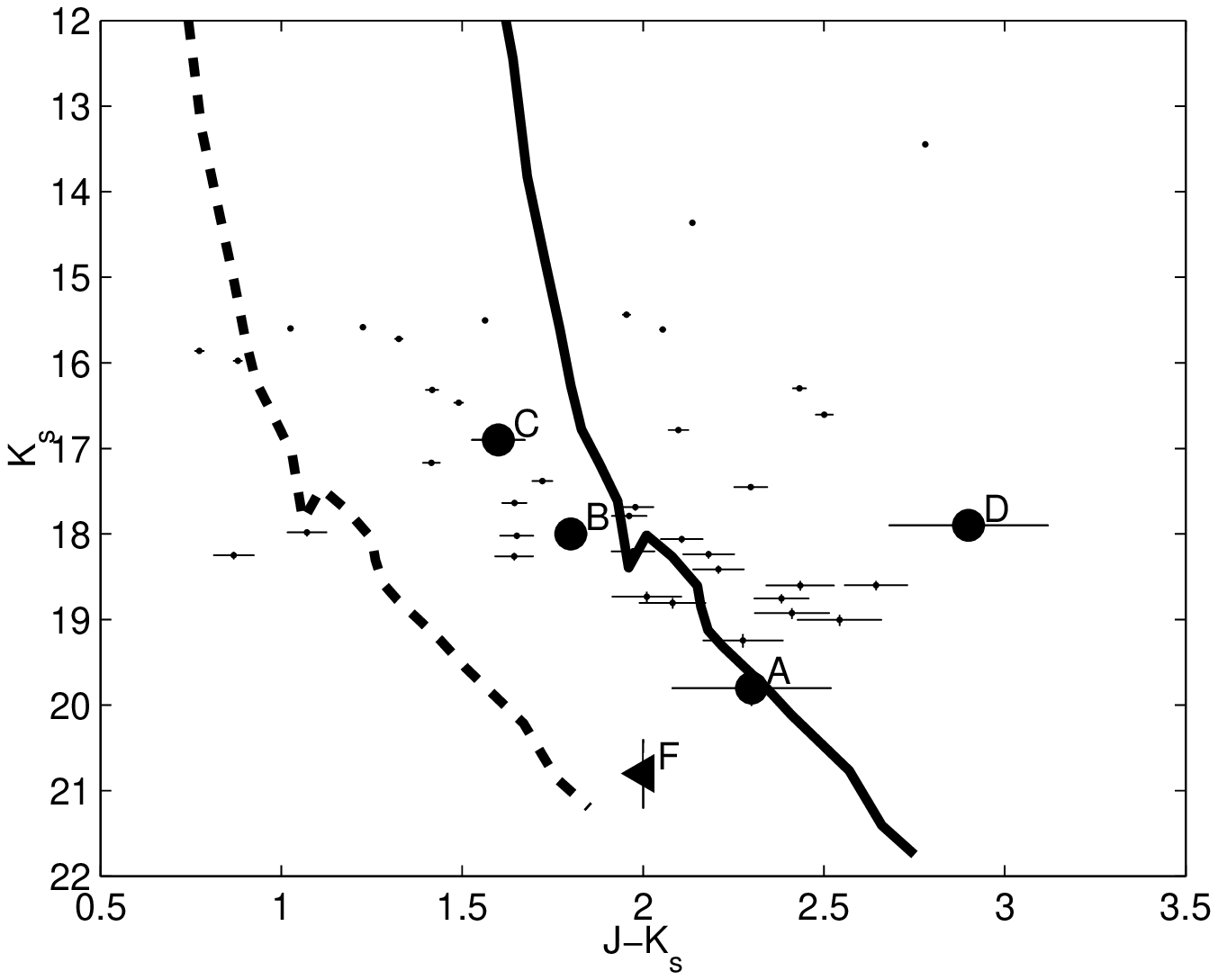}
\caption{Color-magnitude diagrams for the sources in Table~\ref{tab:stars}
(circles) and 37 field stars (points).  Also plotted are model
main-sequences, assuming for $A_{V}=10$~mag (solid line) and $A_{V}=5$~mag
(dashed line), both at a distance of 8~kpc.  $J$ vs.\ $m_{\rm 50CCD}-J$ (left); $K_{s}$
vs.\ $J-K_{s}$ (right).} 
\label{fig:hr}
\end{figure*}

In Figure~\ref{fig:cc}, we show a
color-color diagram for the sources in Table~\ref{tab:stars} and 37
field stars for which we were able to perform reasonably accurate
photometry in all bands.  The sources from Table~\ref{tab:stars} all
lie in the region populated by the other sources.  In Figure~\ref{fig:cc}
we also show model main-sequence colors \citep[based on][p.\ 388]{allen} for $A_{V}=5$--10~mag.
These fit the majority of the points, with the exception of source D.
However, since source D was blended with source C (making the IR
photometry less reliable) and it  is still in the region of the
color-color diagram populated by other stars (some of whom were
slightly blended themselves), we consider it unlikely
that D has significantly non-stellar colors, and therefore is probably
not a counterpart to \sgr\ (both the disk models for AXPs [\citealt*{phn00}]
and the optical/IR detection of an AXP [\citealt*{hvkk00}] have
non-stellar colors, and we expect something similar for SGRs).  
In Figure~\ref{fig:hr} we show color-magnitude diagrams for the
sources in Table~\ref{tab:stars} and 37 field stars.  Again, the
sources populate the same region (although the field stars were chosen
to be somewhat brighter than the faintest of the candidate sources),
and are well bounded by model main sequences with $d=8$~kpc and
$A_{V}=5$--10~mag.  Source D, in particular, has slightly strange
colors (also as seen in Figure~\ref{fig:cc}) but its magnitudes are
well within the range of the main sequence.  Main sequences with
$d=12$--15~kpc and $A_{V}\approx 
19.2$~mag, such as those plotted in \citet{vhl+00}, are too faint and
too red to match these sources.  Therefore the sources in
Table~\ref{tab:stars} are likely to be in the 
foreground of the cluster found by \citet{vhl+00}.

\begin{figure*}
\plotone{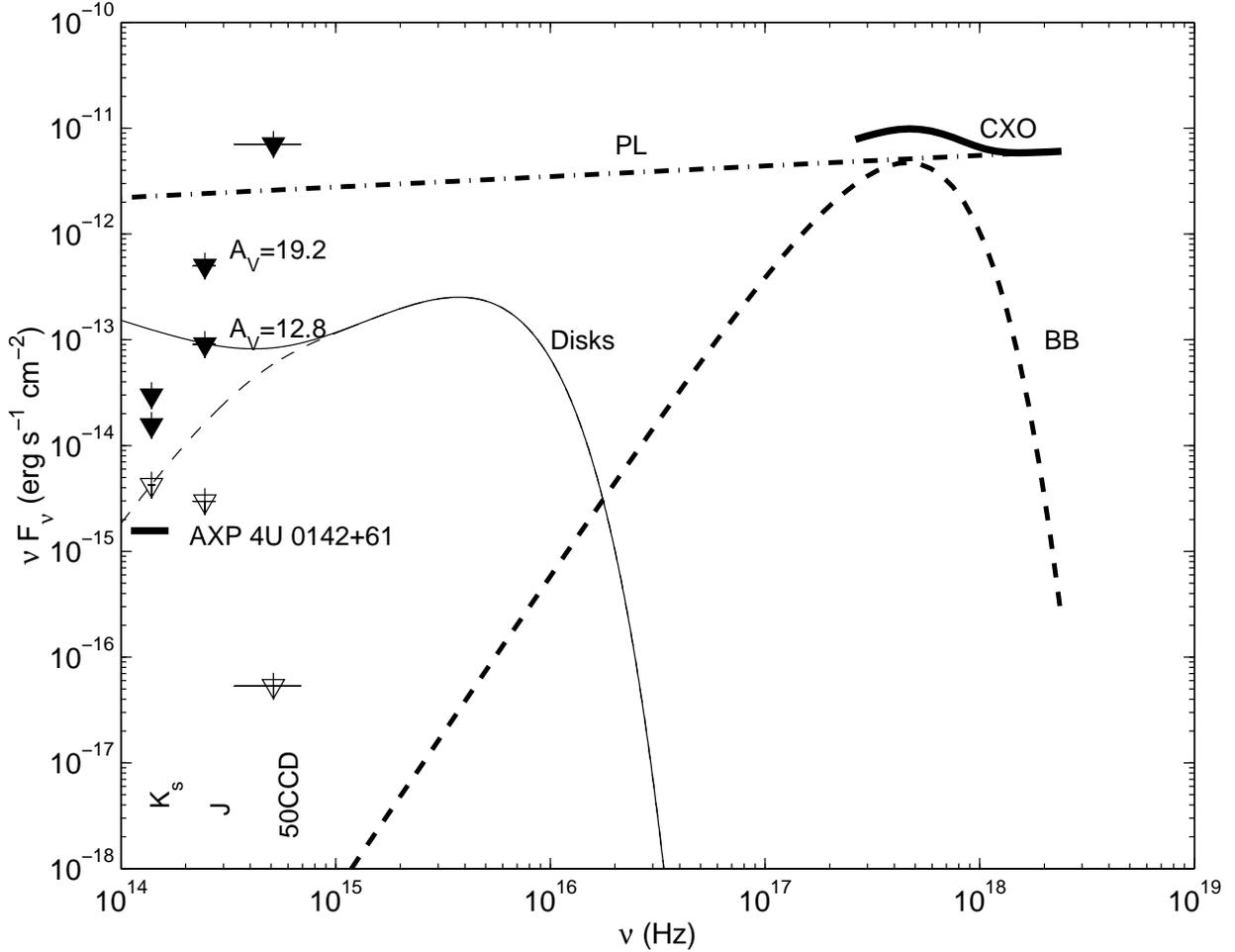}
\caption{Spectral energy distribution for \sgr, including {\em CXO}\
power-law + blackbody spectrum (\citealt{fkkf01}; also see
\citealt{hlk+99}) and upper limits.  Limits are raw values 
(open triangles), and corrected for extinction with $A_{V}=12.8$~mag
and $A_{V}=19.2$~mag (filled triangles).  
Solid thick line marked ``CXO'' is the measured {\em Chandra} spectrum;
the thick dotted line marked ``BB'' is the blackbody component of the
fit, and the thick dot-dashed line
marked ``PL'' is the power-law component, both extrapolated to lower energies.
The thick line marked ``AXP
4U~0142+61'' is the extinction corrected  detected $K_{s}$ flux of 4U~0142$+$61
(F.\ Hulleman 2001, private communication) scaled by the ratio of
X-ray fluxes of 4U~0142$+$61 and \sgr.  Also plotted are the basic 
disk model (thin solid line) from \citet{phn00}, which the current limits
rule out, and a disk truncated at $10^{10}$~cm (thin dashed line).} 
\label{fig:nufnu}
\end{figure*}

Source G in Figures~\ref{fig:stis} and
\ref{fig:nirc} and
in Table~\ref{tab:stars} is too faint for accurate photometry
given its proximity to 
other sources.  It could  be a counterpart of
\sgr, and we require additional deep, high-resolution infrared imaging to
settle this issue.
Regardless, it is  near our detection limit, so we can place
general upper limits to 
the emission from \sgr\ of $m_{\rm 50CCD} \gsim \stislim$~mag,
$J\gsim\jlim$~mag, and $K_{s}\gsim \klim$~mag, consistent with
previously published limits  \citep{ed00,vhl+00}.  We show these
limits along with the {\em CXO}\ X-ray spectrum \citep{fkkf01} in
Figure~\ref{fig:nufnu}.  These limits translate into unabsorbed (``U'') 
X-ray-to-infrared flux ratios $f^{\rm U}_{X}/(\nu_{J} f_{\nu,J}^{\rm
U}) \gsim 160$ and  $f^{\rm U}_{X}/(\nu_{K_{s}} f_{\nu,K_{s}}^{\rm
U}) \gsim \expnt{1.1}{3}$ (assuming $A_{V}=12.8$~mag).  This last limit is only a factor of $\sim 10$
less than  the detected
X-ray-to-$K_{s}$ limit found for the AXP 4U~0142$+$61 (F.\ Hulleman 2001,
personal communication; see Figure~\ref{fig:nufnu})
and thus we can exclude the basic disk
model for AXP/SGR emission (\citealt{phn00}; see also
\citealt{kkvk+01}).  If a disk does exist, then it must be
truncated at 
$R\approx 10^{10}$~cm.  As seen for SGR~0526$-$66 \citep{kkvk+01}, the
optical/IR limits require that the X-ray power-law break between the 
X-ray and the optical bands, but allow for a continuation of the X-ray
blackbody. 

We were  able to detect but not photometer many luminous members of
the massive star cluster that \citet{vhl+00} proposed as the origin of \sgr, as 
we only had a small number of exposures where the central M giant stars were
unsaturated.  In the remaining exposures these stars were too saturated
to allow good psf subtraction, and their scattered light, bleed
trails, and readout trails prevent accurate measurement  of the stars
in the cluster.

\subsection{Radio}
We do not detect either \sgr\ or \psr\ in the radio images.  We
therefore have 3-$\sigma$ limits on point source emission of 6~mJy
(332~MHz) and 1.5~mJy (1.4~GHz).
We also have 3-$\sigma$ limits on extended emission of $6.1\mbox{ mJy
arcmin}^{-2}$ (332~MHz) and $6.2\mbox{ mJy arcmin}^{-2}$ (1.4~GHz).

We were able to detect SNRs \snr, G043.3$-$00.2, G041.1$-$00.3, and
possibly G043.9$-$01.6 at 332~MHz (Figure~\ref{fig:p-wide}). However,
 G041.1$-$00.3, G043.3$-$00.2, 
and G043.9$-$01.6 are too attenuated by the primary beam of
the VLA antennas for the flux measurements to be useful.  For \snr,
we measure $S_{332\smhz}=1.24 \pm 0.08$~Jy, excluding the
contributions  of point sources (see Figure~\ref{fig:p-zoom}), but we
are likely missing some flux on the shortest baselines so this flux may be
 underestimated by 5\%--10\%.

\subsubsection{Serendipitous Sources}
\label{sec:snr}
\begin{figure*}
\plotone{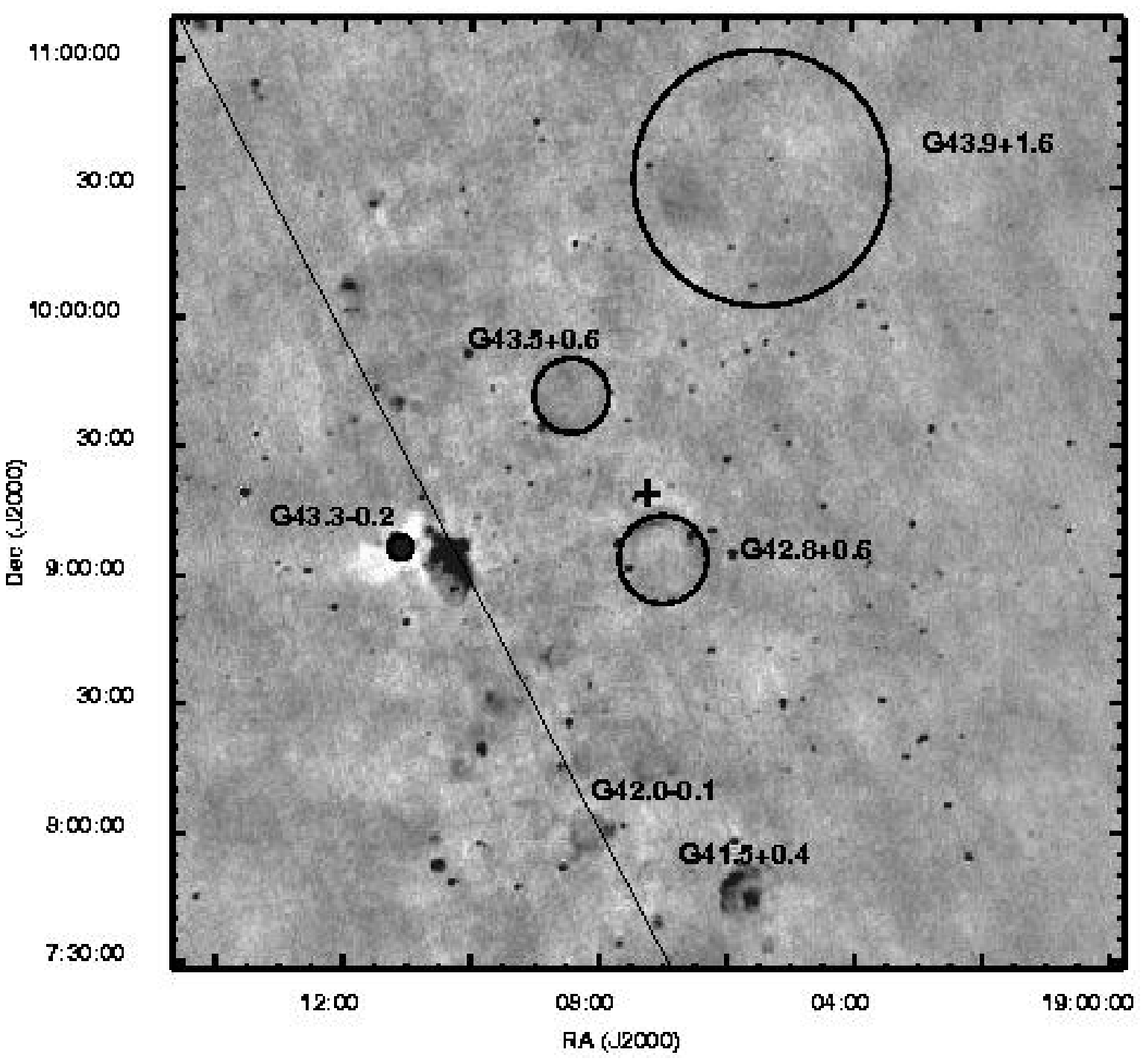}
\caption{Wide-field 332-MHz VLA image of the region around \sgr.  SNRs are marked by circles
and/or labeled.  The diagonal line is the Galactic
plane.  The cross
marks the position of \sgr.  \nsnr, G042.0$-$00.1, and G041.5+00.4  are
new candidate SNRs;  see also
Section~\ref{sec:snr} and Figures~\ref{fig:snr} and \ref{fig:p-hii}.  There 
are a number of smaller \ion{H}{2} regions that we have not labeled.
The horizontal and
vertical lines visible in some places are artifacts created when the
planar facets were combined (see Section~\ref{sec:radio}).
\label{fig:p-wide}
}
\end{figure*}

The 332-MHz image (Figure~\ref{fig:p-wide}) shows three
 sources besides the known SNRs that have shell-like morphologies
but are in neither the current version of the Galactic SNR catalog
 \citep{g00} nor  any online
database.  We consider these sources to be candidate  SNRs, but 
with only limited data at other wavelengths we are unable to confirm
 an SNR classification for any of the three.  The
sources could be \ion{H}{2} regions or other similar sources and
require additional confirmation.  Here we  present what limited
 information we have on each.

\begin{figure*}
\plotone{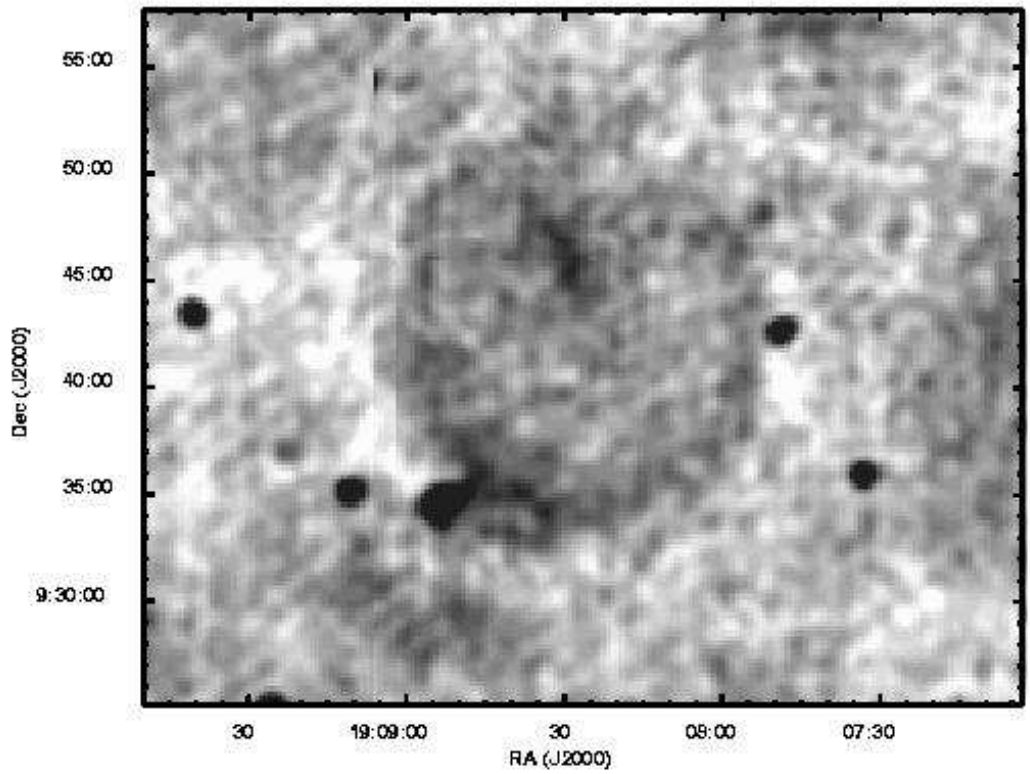}
\caption{332-MHz VLA image of  candidate SNR \nsnr.\label{fig:snr}}
\end{figure*}

The first source is \nsnr\ (see Figure~\ref{fig:snr}).
Its center is at  position $\alpha_{2000}=19^{\rm h}08^{\rm m}26^{\rm
s}$, $\delta_{2000}=+09\degr42.0\arcmin$.  It is circular, with radius $\approx
8.7\arcmin$.  The total flux at 332~MHz is $1.05\pm 0.08$~Jy
(excluding the point source to the south-east), with an average
surface brightness of $\approx 3.9\mbox{ mJy arcmin}^{-2}$.  The edge
appears to be brighter, with typical surface brightness of 6--8$\mbox{ mJy
arcmin}^{-2}$, while the interior is more typically 1--3$\mbox{ mJy
arcmin}^{-2}$.  

Without a measurement at another frequency, it is hard to determine
the properties of \nsnr.  We were not able to find any archival X-ray or radio data at this
position that could definitively classify \nsnr.  However, there is a
marginal detection in the 1.4-GHz NRAO VLA Sky Survey
\citep[NVSS;][]{ccg+98}.  The source is present in the NVSS with the same
general morphology although it is too faint to accurately measure a flux.  This
faintness, though, suggests that \nsnr\ is not a thermal source, as it
would have a NVSS flux of $\sim 1$~Jy and would then be 
easily detectable.
There are two X-ray sources from the {\em EXOSAT} medium-energy slew
survey catalog \citep{rph+99} that lie in or near \nsnr:
EXMS~B1906+095 and EXMS~B1906+094.  Both sources had been
associated  with the X-ray binary 4U~1907+097, which is within their
$\sim 10\arcmin$ position uncertainties, but  the
positions of the {\em EXOSAT} sources are closer to \nsnr\ than to
4U~1907+097 \citep{sgb+80}.  It is therefore conceivable that the
\textit{EXOSAT} sources are associated with regions of \nsnr, but we
require additional X-ray imaging to confirm this.

We can estimate the distance to \nsnr\ very
roughly using the updated $\Sigma$-$D$ relation \citep{cb98}.
Assuming a spectral index $\alpha=0.5$, appropriate for many
shell-type SNRs \citep[e.g.][]{cb98}, we find
$\Sigma_{1\sghz}=\expnt{3.8}{-21}\mbox{ W m}^{-2}\mbox{ Hz}^{-1}\mbox{
ster}^{-1}$, which implies a linear diameter $D\approx 40$~pc and
distance $d\approx 8.1$~kpc.  We note that these numbers are extremely
imprecise (given both intrinsic uncertainties in the $\Sigma$-$D$
relation and in estimates of the 1-GHz surface brightness) probably
having  errors approaching 100\%.  However, this 
value is consistent with \nsnr\ being near the tangent point of spiral
arm 3 of the  \citet{tc93} electron density model.

\begin{figure*}
\plotone{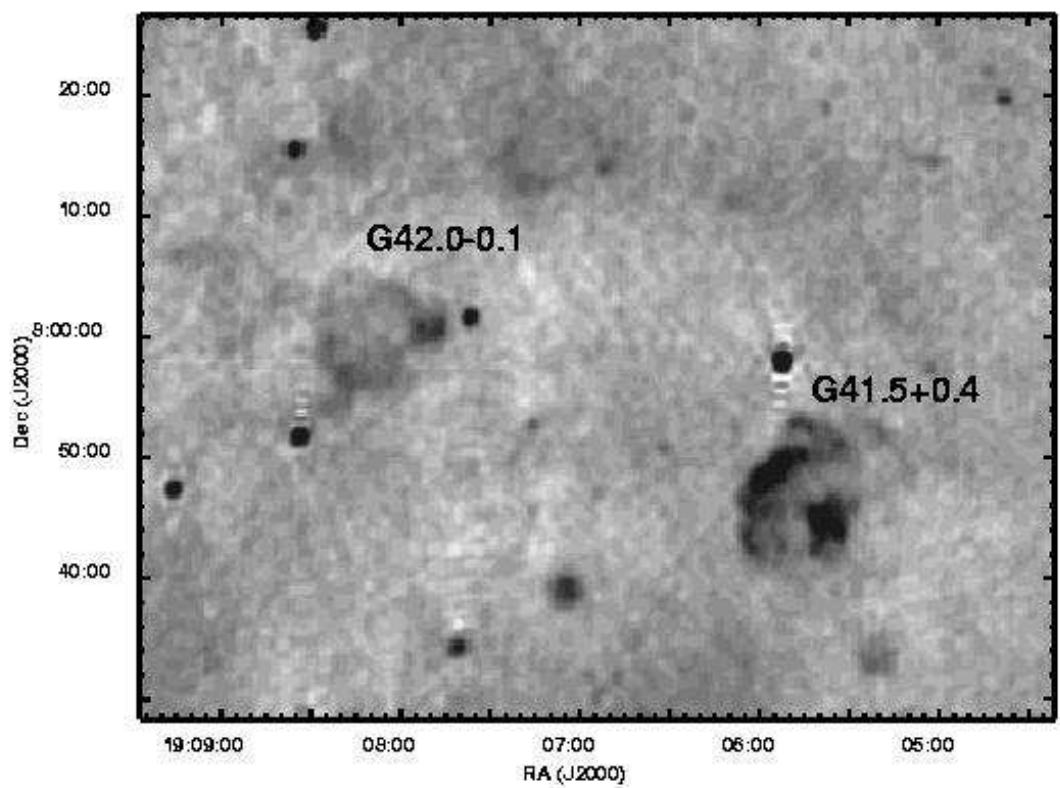}
\caption{332-MHz VLA image of  candidate SNRs G042.0$-$00.1 
and G041.5+00.4.\label{fig:p-hii}}
\end{figure*}

Also intriguing are the objects in Figure~\ref{fig:p-wide} marked
G042.0$-$00.1 and G041.5+00.4 (see Figure~\ref{fig:p-hii}).  These
objects have morphologies that are 
less well-determined than that of \nsnr\ and are further out in the
VLA primary beam, so we cannot make very useful measurements, but their
shapes are suggestive of either  SNRs or \ion{H}{2} regions.
G042.0$-$00.1 is centered at  $\alpha_{2000}=19^{\rm h}08^{\rm m}11^{\rm
s}$, $\delta_{2000} =+08\degr00.5\arcmin$, is roughly circular with radius
$4.3\arcmin$, and has a 332-MHz flux of $1.8 \pm 
0.4$~Jy (this value has been corrected for the attenuation of
the primary beam of the VLA antennas and is therefore highly uncertain).
The protruding region to the 
south-east of G042.0$-$00.1 was identified as an \ion{H}{2} region on the basis
of radio recombination lines by \citet*{lph96}, but their data do not
mention the rest of the complex and their angular resolution should be able to
distinguish between the center of G042.0$-$00.1 and the smaller region.
G042.0$-$00.1 is present in the NVSS with similar morphology to the
332-MHz image and flux
$S_{\rm NVSS} = 0.3 \pm 0.1$~Jy, suggesting a non-thermal spectrum and
leading to a provisional classification as an SNR.

G041.5+00.4 is centered at  $\alpha_{2000}=19^{\rm h}05^{\rm m}
48^{\rm s}$, $\delta_{2000}=+07\degr46.1\arcmin$.  It has a complicated
morphology, with a brightened rim to the north-east and a compact core
to the south-west ($\alpha_{2000}=19^{\rm h}05^{\rm m}37^{\rm s}$,
$\delta_{2000}=+07\degr44.8\arcmin$), suggestive of a
pulsar wind nebula.  The source has an overall radius of $4.9\arcmin$
and a 332-MHz flux of $1 \pm 1$~Jy (again, this is a highly uncertain
value). 
Comparison of the overall 332-MHz flux with the NVSS flux of $1.3 \pm
0.2$~Jy leads to a flat 
spectrum like that of core-dominated SNR or an \ion{H}{2} region.
We need more data to conclusively classify all three of these objects;
further analysis and observations are in  progress.

The wealth of information available in 
Figure~\ref{fig:p-wide},  concerning both  already known objects and newly
discovered ones, is readily apparent.  It 
serves to illustrate how productive low-frequency wide-field imaging
of the Galactic plane with high resolution can be \citep[e.g.][]{lkl+00}.

\acknowledgements
We would like to thank J.~Lazio for his assistance with the
low-frequency data reduction and D.~Fox for useful comments.
DLK is supported by the Fannie and John Hertz Foundation, SRK by 
NSF and NASA, and MHvK by a fellowship from the Royal Netherlands
Academy of Arts and Sciences.     
The NASA/ESA {\em Hubble Space Telescope}\ is operated by the Association of
Universities for Research in Astronomy, Inc. under NASA contract
No. NAS5-26555.The VLA is
operated by the National Radio Astronomy 
Observatory, which is a facility of the National Science Foundation
operated under cooperative agreement by Associated Universities, Inc.
The W.M. 
Keck Observatory is operated as a scientific partnership among
the California Institute of Technology, the University of California
and the National Aeronautics and Space Administration.  The
Observatory was made possible by the generous financial support of the
W.M. Keck Foundation.  
We have made extensive use of the SIMBAD database, and we would like
to express our appreciation of the astronomers who maintain this database.

\bibliographystyle{apj}

\begin{deluxetable}{l l l r r l}
\tablecaption{Observation Summary\label{tab_obs}}
\tablecolumns{6}
\tablehead{
\colhead{Date} & \colhead{Telescope /} & \colhead{Band} &
\colhead{$\lambda$} & \colhead{Exposure}
& \colhead{Comments} \\
 & \colhead{Instrument} & & & 
 \colhead{(min)} & \\}
\startdata
1998~Sep~05 & Keck~I/NIRC & $K_{s}$ & 2.15 $\um$ & 49.7 &
 $0\farcs35$ seeing\\
 & & $J$ & 1.25 $\um$ & 22.5 & \\
1998~Nov~18 & VLA & L & 20~cm & 76 & D-config. \\
1998~Dec~26 & VLA & P & 90~cm & 441.5 & C-config., 32 channels\\
1999~Apr~26 & VLA & L & 20~cm & 116 & C-config.\\
2000~Apr~12 & {\em HST}/STIS & 50CCD & 0.585 $\um$ & 86.6 &
\\
\enddata
\end{deluxetable}

\begin{deluxetable}{c l l l l l}
\tablecaption{Stars Near \sgr\ Error Circle\label{tab:stars}}
\tablewidth{34pc}
\tablehead{
\colhead{Label} & \mc{2}{c}{Position\tablenotemark{a}\ (J2000)} & \mc{3}{c}{Magnitude} \\
 & \colhead{$\alpha$} & \colhead{$\delta$} &
 \colhead{${\rm 50CCD}$} & \colhead{$J$} & \colhead{$K_{s}$} \\

}
\startdata
A & $19^{\rm h}07^{\rm m}14\fs41$ & $+09\degr19\arcmin19\farcs2$ &
$28.1\pm 0.2$ & $22.1 \pm 0.2$ & $19.8 \pm 0.2$ \\		     
B & $19^{\rm h}07^{\rm m}14\fs29$ & $+09\degr19\arcmin19\farcs0$ &
$26.16\pm 0.06$ & $19.8\pm 0.04$ & $18.0 \pm 0.1$\\		     
C & $19^{\rm h}07^{\rm m}14\fs31$ & $+09\degr19\arcmin19\farcs8$ &
$25.02\pm 0.05$ & $18.5 \pm 0.07$ & $16.9 \pm 0.1$  \\		     
D & $19^{\rm h}07^{\rm m}14\fs29$ & $+09\degr19\arcmin20\farcs0$ &
$26.46\pm 0.06$ & $\klim \pm 0.2$ & $17.9 \pm 0.1$\\
E & $19^{\rm h}07^{\rm m}14\fs40$ & $+09\degr19\arcmin22\farcs2$ & 
$> \stislim$ & $> \jlim$ & $20.6 \pm 0.4$ \\
F & $19^{\rm h}07^{\rm m}14\fs38$ & $+09\degr19\arcmin18\farcs9$ &
$> \stislim$ & $> \jlim$ & $\klim \pm 0.4$ \\ 
G & $19^{\rm h}07^{\rm m}14\fs34$ & $+09\degr19\arcmin20\farcs1$ & 
$> \stislim$ & $\gsim \jlim$ & \nodata \\
\enddata
\tablecomments{See Figures~\ref{fig:stis} and \ref{fig:nirc}.}
\tablenotetext{a}{Accurate to $\pm 0\farcs21$.}
\end{deluxetable}

\end{document}